\documentclass{aa}
\usepackage{graphicx}
\begin{document}

%\thesaurus{ 03.13.5; 04.19.1; 10.07.1; 13.18.2  }

\title{A radio continuum survey of the southern sky at 1420 MHz}
\subtitle{Observations and data reduction}

\author{J.C. Testori \inst{1}, P. Reich  \inst{2},
J.A. Bava \inst{1}, F. R. Colomb \inst{1}, E.E. Hurrel \inst{1},
J.J. Larrarte \inst{1}, W. Reich \inst{2}, \\ and A.J. Sanz \inst{1}
}
\offprints{P. Reich}
\institute{Instituto Argentino de Radioastronom\'{\i}a, C.C. 5,
      1894 Villa Elisa (Prov. de Bs.As.), Argentina
  \and Max-Planck-Institut f\"ur Radioastronomie, Auf dem
     H\"ugel 69, D--53121 Bonn, Germany
}
\date{Received 7 November 2000 / Accepted 15 January 2001}
\abstract{
We describe the equipment, observational method and reduction procedure of
an absolutely calibrated radio continuum survey of the South Celestial
Hemisphere at a frequency of 1420~MHz. These observations cover the
area $\rm 0^h \leq R.A. \leq 24^h$ for declinations less than
$-10\degr$. The sensitivity is about 50~mK T$_{\rm B}$ (full beam
brightness) and the angular resolution (HPBW) is  $35\farcm4$, which
matches the existing northern sky survey at the same frequency.
}

\authorrunning{J.C. Testori et al.}
\titlerunning{Southern sky at 1420 MHz}

\maketitle

\section{Introduction}

The present paper describes the observations of the South Celestial
Hemisphere at a frequency of 1420~MHz carried out with one of the 30-m
radio telescopes of the Instituto Argentino de Radioastronom\'{\i}a
(IAR) at Villa Elisa, Argentina. This continuum survey is intended to
complement the northern sky survey (Reich\  \cite{reich82}; Reich \&
Reich\ \cite{reich+reich86}) made with the 25-m Stockert telescope near
Bonn to an all-sky radio continuum survey at 1420~MHz.

There are many reasons for carrying out such a survey. Only
all-sky surveys allow, under reasonable assumptions, modelling of
the emission distribution in the Galaxy. The recognition and study of
nearby large-scale features needs survey data as well as investigations
of the Galactic emission spectrum when comparing well-calibrated survey
data at different frequencies. One important aspect is the
determination of the separation of thermal and non-thermal emission
in addition to the non-thermal spectral index, which varies with
frequency. A spectral index map based on the 408-MHz survey of Haslam
et al. (\cite{haslam+82}) and the northern sky 1420-MHz survey by Reich
\& Reich (\cite{reich+reich88a}) showed rather unexpected flat spectra
towards the anti-centre direction. In a  subsequent discussion
Reich \& Reich (\cite{reich+reich88b}) proposed an explanation in
terms of a cooling--convection halo model. Spectral information for the
southern sky is needed to extend and refine this modelling. Galactic
foreground emission components and their spectra are also of high
interest for cosmic microwave background studies. Meanwhile, a number
of sky-horn measurements were available up to short cm-wavelength (e.g.
Platania et al. (\cite{platania+98}) and references therein), which
give fairly consistent spectral data for the large-scale Galactic
emission. However, due to the rather coarse angular resolution of
sky-horns, no spatial details can be derived. These require data of
higher angular resolution that are provided by single-dish telescopes.

In the following sections we describe the receiver (2), the
observation procedure (3), the data acquisition system (4) and the
data processing (5). The survey maps will be presented in a forthcoming
paper. As examples we show the centre and the anti-centre region in
comparison with the northern sky survey to demonstrate its similar
performance.

\section{Receiving equipment}

A deep all-sky radio continuum survey requires receiving equipment
having not only sufficient sensitivity to reach the confusion
limit of the telescope, but also with the ability to measure the
brightest regions of the sky. This is particularly important when
regions of both low and high brightness will be crossed by the same
scan. The high scanning speed ($10\degr$/min) necessary to perform a
large-area survey makes it impracticable to attenuate the signal from
brighter objects during the observations. Thus, a receiver should
possess a linear dynamic range in excess of 40~dB. In addition, many
problems can be avoided if the receiver calibration is constant over
long periods of time, even in storage or standby, for observing
periods spaced at intervals of several months. The extra requirements
of night-time observations avoiding solar and terrestrial interference
make telescope scheduling difficult at times.

% Fig. 1
\begin{figure}
\includegraphics[width=\hsize]{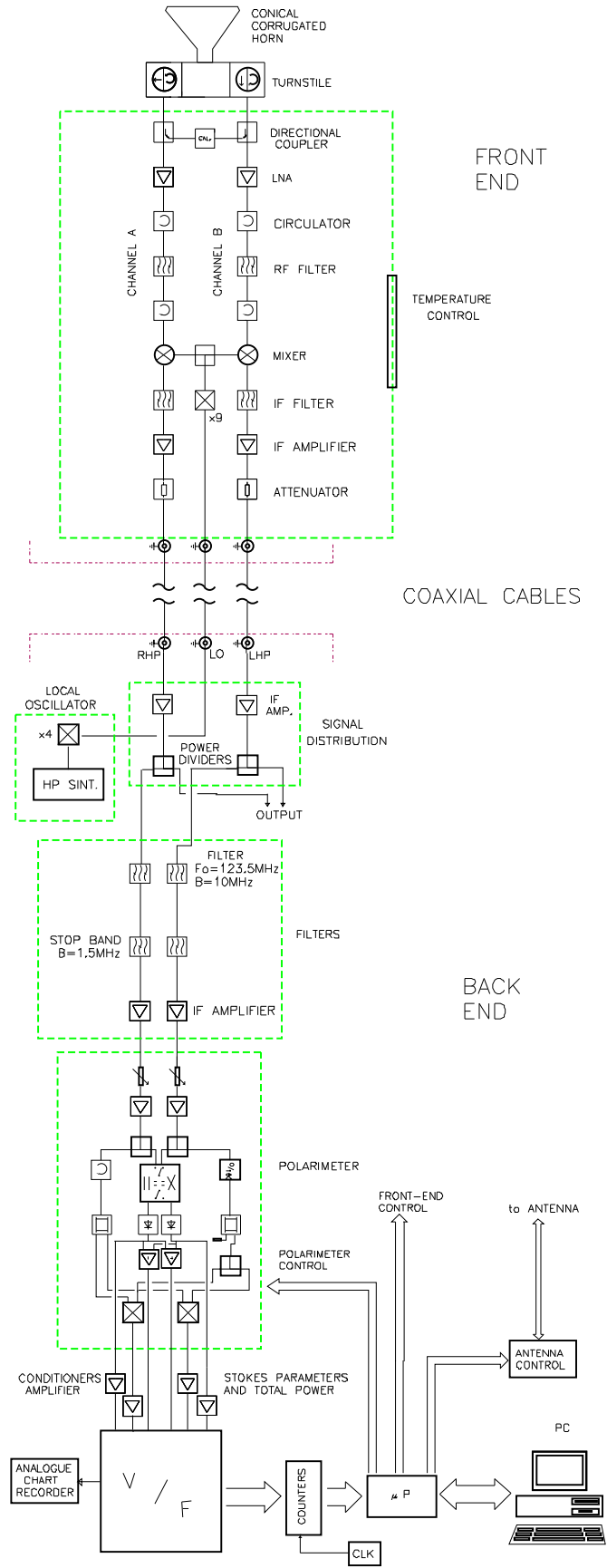}
\caption{Block diagram of the 1420-MHz continuum receiver at Villa Elisa}
\label{fig_one}
\end{figure}

A receiver satisfying these requirements was developed and installed in
the prime focus of one of the 30-m telescopes of the IAR.
Figure~\ref{fig_one} shows the block diagram of the receiver system.

The feed is a corrugated horn which, at 1420~MHz, illuminates the
dish with a half power beam width of $60\degr$. Its attenuation at
the edge of the dish is $-$17~dB. The feed is followed by a
`turnstile', a passive device which picks up two linear or circular
polarization components, depending on the adjustment. In our case the
adjustment was made such as to allow the extraction of the left-hand
and right-hand circular polarization of the signal with a polarization
isolation larger than 30~dB. The connection between the feed and the
turnstile is  made by means of a circular waveguide and the two
polarization components, extracted from the turnstile, are coupled out
by a coaxial probe.

The two circular polarization components are amplified in two separate
channels, each of which contains a GaAs FET low-noise amplifier (LNA)
with an equivalent noise temperature of 60~K at room  temperature. The
system noise temperature, including the receiver, ground and sky
contributions, is 90~K towards the coldest regions of the sky. Both
channels of the receiver were calibrated by measuring the intensity of
a standard noise diode, injected through a pair of directional couplers
connected between the turnstile and the LNAs, with two optional
levels of either 10~K or 50~K. Control signals from the control room can
switch the noise source and select its level. The rest of the
front-end receiver consists of bipolar amplifiers, interdigital
filters, mixers, a local oscillator and an IF-stage with a centre
frequency at 123.5~MHz. The IF-signals from the front-end were
subsequently filtered with a phase-matched pair of filters and fed into
an IF-polarimeter. This device was supplied by the MPIfR, Bonn, where
similar polarimeter systems are used at the Effelsberg 100-m
telescope (e.g. Schmidt \& Zinz\ \cite{schmidt+zinz90}). The output is
proportional to the four Stokes parameters I, V, U and Q. The linear
polarization parameters U and Q are obtained from the in-phase and
quadrature analogue correlation of the left-hand and right-hand
circular signals coming out from the channels A and B of the receiver.
The remaining two parameters, I and V, are formed from the sum and
difference of the detected input signals. The receiver front-end is
thermally controlled to a temperature of $20\degr \pm 2\degr$~C which
results in a gain stability of about 0.1~dB for the system during four
hours of observations. Special care was taken to match the cable
lengths from the noise source to the directional couplers and from the
local oscillator to the mixers. Also, in order to minimize the phase
drift in the polarimeter due to differential changes in electrical
lengths of channels A and B and to changes in temperature, the
underground cables used for IF and LO between the front-end and the
control room have low attenuation (Heliax type) and high phase
stability. In addition, periodic phase calibrations were made to
ensure that the phase tracking between channels was less than 5\degr.
In Table~\ref{tab_one} the specifications of the receiver are
summarized.

%Table1
\begin{table}
\caption[ ]{Receiver specifications}
\label{tab_one}
\begin{tabular}{ll}
\noalign{\smallskip}
\hline
\noalign{\smallskip}
RF-IF gain                     & 80 dB\\
Line emission rejection        & $\geq 23$~dB\\
Intermediate frequency         & 123.5 MHz\\
Dynamic range                  & 40 dB\\
Polarization components        & Circular\\
Polarization isolation         & $\geq 30$~dB\\
System temperature             & $\sim 90$~K\\
Gain stability in four hours   & 0.1 dB\\
Illumination at the edges of the 30-m \\
\quad reflector                &$-$17 dB\\
Side lobe level related to the main lobe  & $\leq 25$~dB\\
Back side lobe level           & $\leq 30$~dB\\
Temperature stability within the box & $\leq 2\degr$~C/hour\\
Phase tracking between channels & $\leq 5\degr$\\
\noalign{\smallskip}
\hline
\end{tabular}
\end{table}

The four-phase switching system is similar to the one used by Haslam
et al. (\cite{haslam+74}) to perform a 408-MHz all-sky survey. With
this scheme the receiver front-end cycles through four phases of
4 x 60~msec duration. The phases are:

\smallskip

Phase 1: Antenna signal

Phase 2: Antenna signal + Calibration

Phase 3: Antenna signal + Calibration + $180\degr$

\hskip 1.45truecm Phaseshift between LHC and RHC

Phase 4: Antenna signal + $180\degr$ Phaseshift

\smallskip

This particular data acquisition sequence was chosen for optimal
calibration because it permits simultaneous measurement and
reduction of the effects of receiver gain variations, second-order
terms in the correlator and variations of the correlator and
voltage-to-frequency converters. For proper receiver data and antenna
position synchronization and recording, two cycles of four phases were
accumulated. An analogue output and a digital display were available
for monitoring purposes and interference detection.

\section{Observations}

The observations were made in the total-power mode in two sequences.
Between 1987 and 1989, we used a receiver centred at 1435~MHz
avoiding \ion{H}{i} line emission. The effective bandwidth was 14~MHz.
Additional observations were made during a second period in 1993 and
1994. Due to serious problems with interference around 1450~MHz, the
system was tuned at a new centre frequency of 1420~MHz. \ion{H}{i}
line emission was rejected with a band stop filter resulting in an
effective bandwidth of 13~MHz. The major part of the observation was
done during the first period. In both periods the observations were
made exclusively during night time in order to minimize solar and
terrestrial interference. Both periods yielded comparable results.

The data were obtained with the nodding scan technique
(Haslam et al.\ \cite{haslam+74}). The telescope moved continously with
a speed of  $10\degr$/min between a declination of $-10\degr$ and the
southern celestial pole. In azimuth it pointed to the local meridian.
That way, the survey covers 24~hours of right ascension spaced by  one
minute of time with a full sampling for all declinations. The northern
limit of the observations provides an overlap of $9\degr$ in
declination with the northern sky survey, which has a lowest
declination of $-19\degr$. This is essential for the matching of both
surveys and improves the quality of data in this region.

As mentioned in Sect.~2, the Stokes parameter U and Q (corresponding to
linear polarization) have also been recorded, but have not been reduced
so far. However, the polarized intensities at 1.4~GHz at medium
Galactic latitudes show unexpected structures almost uncorrelated with
total intensities (e.g. Uyan{\i}ker et al.\ \cite{uyanik+99}).
Modulation by the interstellar medium in front of highly polarized
background emission is the most likely explanation. Therefore, the
reduction of the polarization data for the southern sky promises
interesting results. We summarize some observational parameters in
Table~\ref{tab_two}.

%Table 2
\begin{table}
\caption[ ]{Observational parameters of the Villa Elisa southern sky survey}
\label{tab_two}
\begin{tabular}{ll}
\hline
\noalign{\smallskip}
Antenna diameter                 & 30 m\\
HPBW (effective)                 & 35\farcm4\\
Aperture efficiency              & 32.8 \% \\
\noalign{\smallskip}\hline
\noalign{\smallskip}
Observing Periods                &\\
1987--1989  (epoch 1)            & \\
~~Centre frequency               & 1435 MHz\\
~~Bandwidth                      & 14 MHz \\
1993--1994 (epoch 2\&3)          &\\
~~Centre frequency               & 1420 MHz \\
~~Effective bandwidth            & 13 MHz \\
\noalign{\smallskip}\hline
\noalign{\smallskip}
Coverage               & $\rm 0^h \leq R.A. \leq 24^h$\\
                       &$-90\degr \leq \delta\leq -10\degr$\\
Sensitivity            & $\sim 50$~mK T$_{\rm B}$\\
                       & (3 $\times$ rms noise)\\
Gain scale accuracy    & 5\% \\
Zero level accuracy    &\\
~~(horn measurements)  & 0.5 K \\
~~(relative to 408-MHz survey) & $\leq 0.2$~K \\
Pointing accuracy      & $\sim 2\arcmin$ \\
S/T$_{\rm B}$ (HPBW=35\farcm4, full beam)  & 11.25 Jy/K \\
S/T$_{\rm A}$ (HPBW=32\arcmin)  & 11.9 Jy/K \\
                                 & (hot-cold-measurement) \\
\noalign{\smallskip}
\hline
\end{tabular}
\end{table}

Radio sources, used as flux and pointing calibrators, were observed
each night. Maps of the calibration sources were obtained in
declination at right ascension intervals of 1 minute. The main
calibration sources used were PKS 0518$-$45 (Pictor~A) and PKS
0915$-$11 (Hydra~A), for which we adopt flux densities of 65.1~Jy and
42.5~Jy, respectively. Other strong radio sources were used as
secondary calibrators. Their fluxes were measured relative to a
main calibrator during one night. Table~\ref{tab_three} lists 22
calibration sources and their adopted peak flux densities.

% Table 3
\begin{table}
\caption[ ]{ Peak flux densities S of the calibration sources}
\label{tab_three}
\begin{tabular}{lr}
\noalign{\smallskip}\hline
\noalign{\smallskip}
Source Name (PKS)  &  S~(Jy) \\
\noalign{\smallskip}\hline
\noalign{\smallskip}
0023$-$26 & 9.1\\
0043$-$42 & 7.5\\
0114$-$21 & 4.1\\
0131$-$36 & 7.1\\
0213$-$13 & 4.4\\
0320$-$37 (For A) & 82.5\\
0453$-$20 & 4.7\\
0518$-$45 (Pic A) & 65.1\\
0741$-$67 & 3.9\\
0814$-$35 & 10.8\\
0915$-$11 (Hyd A) & 42.5\\
1018$-$42 & 4.3\\
1123$-$35 & 2.3\\
1302$-$49 & 7.1\\
1333$-$33 & 12.0\\
1504$-$16 & 3.1\\
1610$-$608 & 60.0\\
1730$-$13 & 6.2\\
1938$-$15 & 6.5\\
2058$-$28 & 5.4\\
2152$-$69 & 32.0\\
2211$-$17 & 8.6\\
\noalign{\smallskip}\hline
\end{tabular}
\end{table}

\section{The data acquisition system}

The data acquisition system used was built around two small
microcomputers: a Commodore~64 (C64) and an IBM~PC connected through a
serial RS 232 channel transmitting in full duplex mode with a speed of
4800 bauds. This system separates signal and calibration, suppresses a
radar signal from the neighbouring international airport of Buenos
Aires (Ezeiza), and provides antenna position acquisition and control.
In addition the separated data streams (approximately 100 bytes per
second) were converted into a tabulated scan array, where a weighted
sum of the data was formed for points separated in declination by
0\fdg25, the `tabular interval', along the length of a scan. The
interpolation was achieved using a tapered `sin(X)/X' interpolation
function whose width was matched to the telescope's beamwidth. During
the time interval between subsequent scans, where the telescope's
direction changes, the tabular scan was computed, formatted and stored
on the IBM PC's disc and printed out on request. Both the computer
hardware and software developed for this survey were adapted from the
observations carried out with the 25-m Stockert telescope for the
northern sky. The subsequent analysis was made by using the NOD2
software library (Haslam\ \cite{haslam74}).

The acquisition module, attached to the on-line processing
microcomputers, consists of two parts: a) signal conditioning
amplifiers and voltage-to-frequency converters (V/F); b) Villa Elisa
Interface (VEI) which connects the C64 with the V/F and the antenna
positioning system. It performed the integration of each 60~msec
receiver phase for each of the four channels and also provided the
electronic timing for the four-phase cycle of the receiver front-end.

The signal conditioning amplifiers provided the necessary amplification
for the required output levels of the polarimeter channels entering the
V/F converter and an offset displacement to achieve the full dynamic
range utilization of the V/F converter. The V/F converter, developed
at the IAR, has an analogue-to-digital conversion of 100~kHz/V,
high linearity and temperature drift stability. Its output is counted
in 15 bit counters.

The VEI is connected with two channels: one is an 8-bit parallel
interface  from the user port of the C64 that allows reading of the
8-byte counter registers of the V/F converter, and the other one is a
serial RS~232 interface built around a Versatile Interface Adapter
(VIA) that interfaces with the antenna positioning system to read the
actual position and to send control commands. In addition, the VEI
provides all the signals necessary to synchronize the real-time
acquisition process: integration time of the analogue-to-digital
converters, reset of the counters, acquisition of Stokes parameter,
antenna position request, noise source injection, phase switching, C64
interrupt and C64 to IBM~PC transmission. The reference clock has
a period of 60~msec, which is the minimum integration time. This value
was chosen to achieve the minimum quantizing error compatible with
receiver parameters, the large range of temperatures to scan and the
maximum full-scale frequency of the  V/F converter. It results in an
1.25$\%$ increase in noise due to digitalization with high 50~Hz
frequency rejection. The error of the signal demultiplexing due to the
curvature of the  gaussian beam  is about 0.44$\%$ of the antenna
temperature for 60~msec integration time.

\section{Data processing}

The data processing is divided into three steps,

\begin{enumerate}
\renewcommand{\labelenumi}{\alph{enumi})}
\item the on-line logging and analysis at the telescope.

\item the off-line analysis to edit and calibrate the data to
form a set of raw maps.

\item combination  of the maps via the NOD-2 program library
(Haslam\ \cite{haslam74}) and absolute calibration
\end{enumerate}

\subsection{The on-line program}

A two-level program was installed on the microcomputers. The first
level was entered every 60~msec via an interrupt given by the VEI to
the C64 and the tasks accomplished during its execution were:

a) Addition of two sets of the four-phase receiver cycles giving
an approximate integration time of 0.5~sec.

b) Recording of the current sidereal time and declination from the
antenna positioning system and computing of both the signal and
calibration demultiplexing for the two total power, the Q and U
channels.

c) Transmission of the recorded and computed values to the IBM~PC
via a 4800 baud RS~232 line.

The second software level provided the weighted tabular scan data.
This IBM~PC program runs at an asynchronous rate and performed
the weighted tabular scan integration at declination intervals of
0\fdg25 and adds start, stop times and scan identification number.
It displays the scan data in real time. A chart recorder permits an
on-line check of both the quality of the astronomical data being
processed and the performance of the total system.

At the end of each scan the transmission between both microcomputers
was suspended to allow the IBM~PC to perform the normalization process,
i.e. a noise calibration average over all the scan calibration data and
storage of the result on disc.

% Fig. 2
\begin{figure}[htbp]
\centering
\includegraphics[width=6.6truecm]{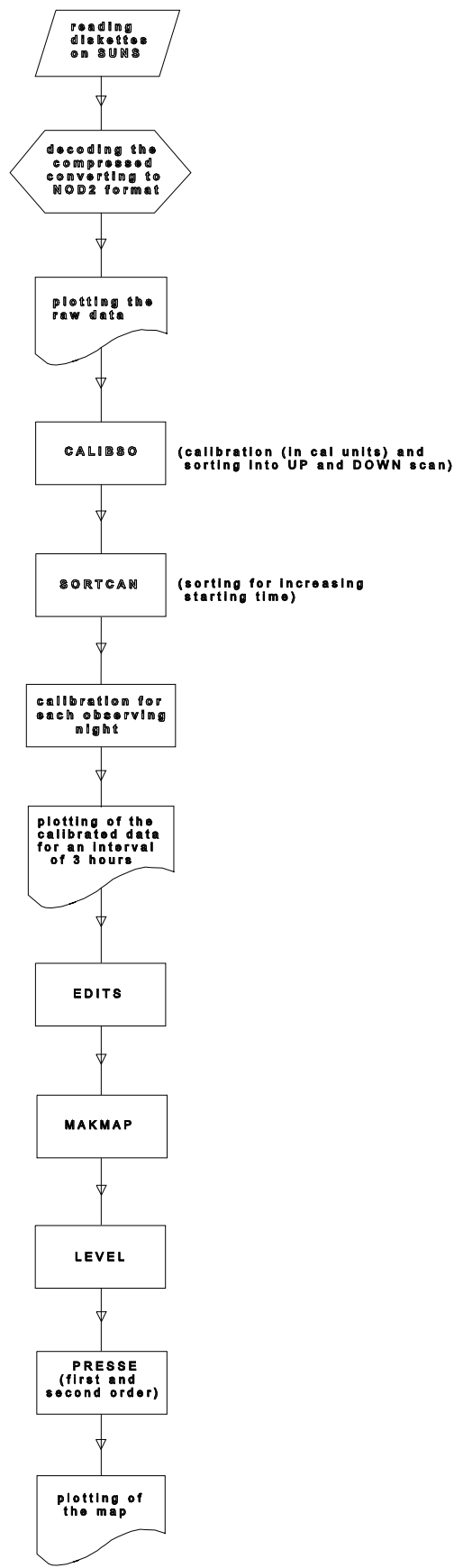}
\caption{Flow diagram of the reduction scheme of the 1420-MHz
continuum survey}
\label{fig_two}
\end{figure}

\subsection{Off-line analysis and map formation}

These reduction steps were performed at the MPIfR, Bonn. A flow diagram
of the off-line analysis software is shown in Fig.~\ref{fig_two}.

An analysis of the raw data, namely TP1, TP2, U and Q and their
corresponding calibration channels, revealed a significantly
reduced quality of the data in one of the TP channels. Hence an
additional coverage in order to reach the required sensitivity was
required.

\begin{figure}[htbp]
\centering
\vbox{
\includegraphics[width=0.70\hsize]{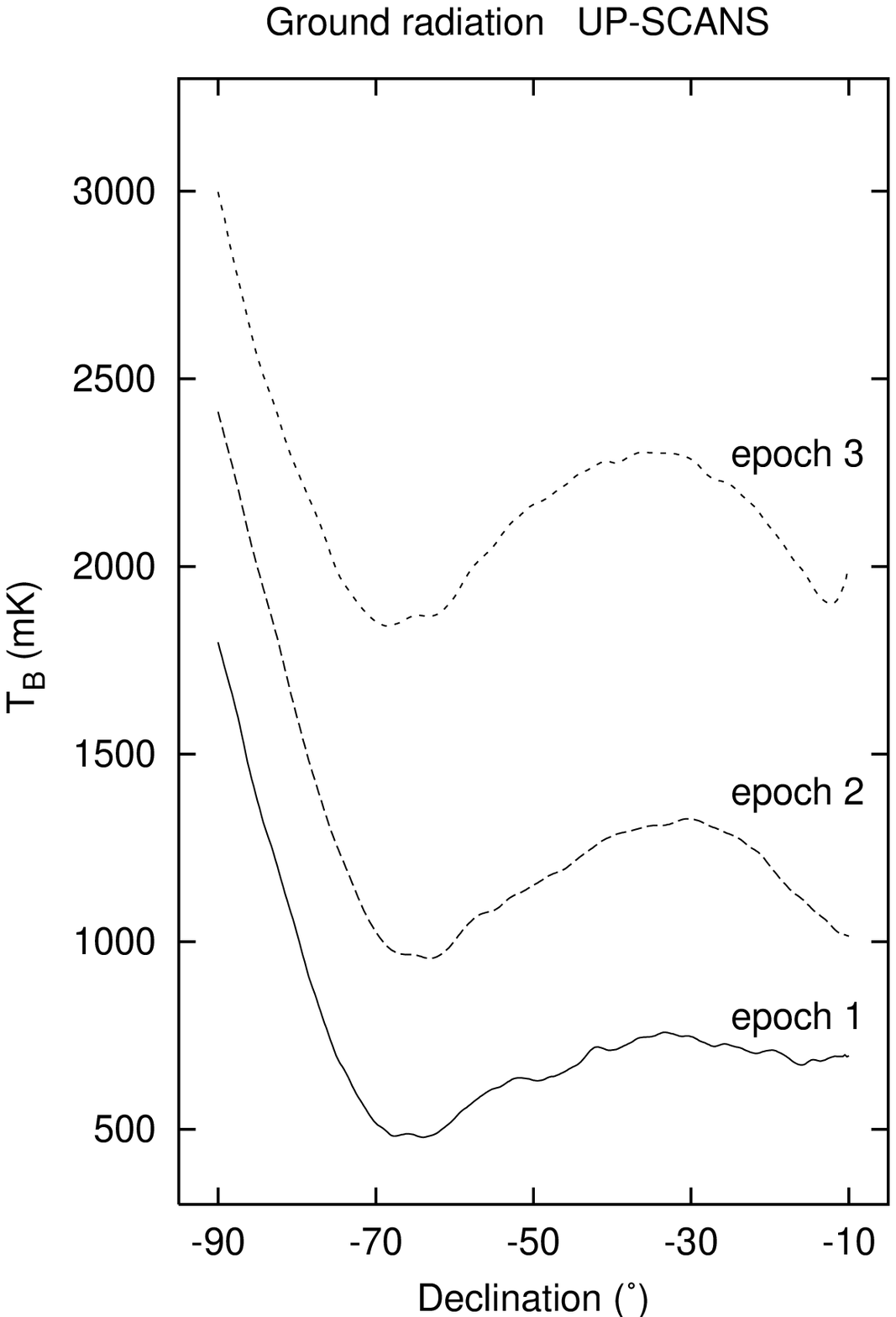}
}
\vbox{
\vspace{0.5cm}
\includegraphics[width=0.70\hsize]{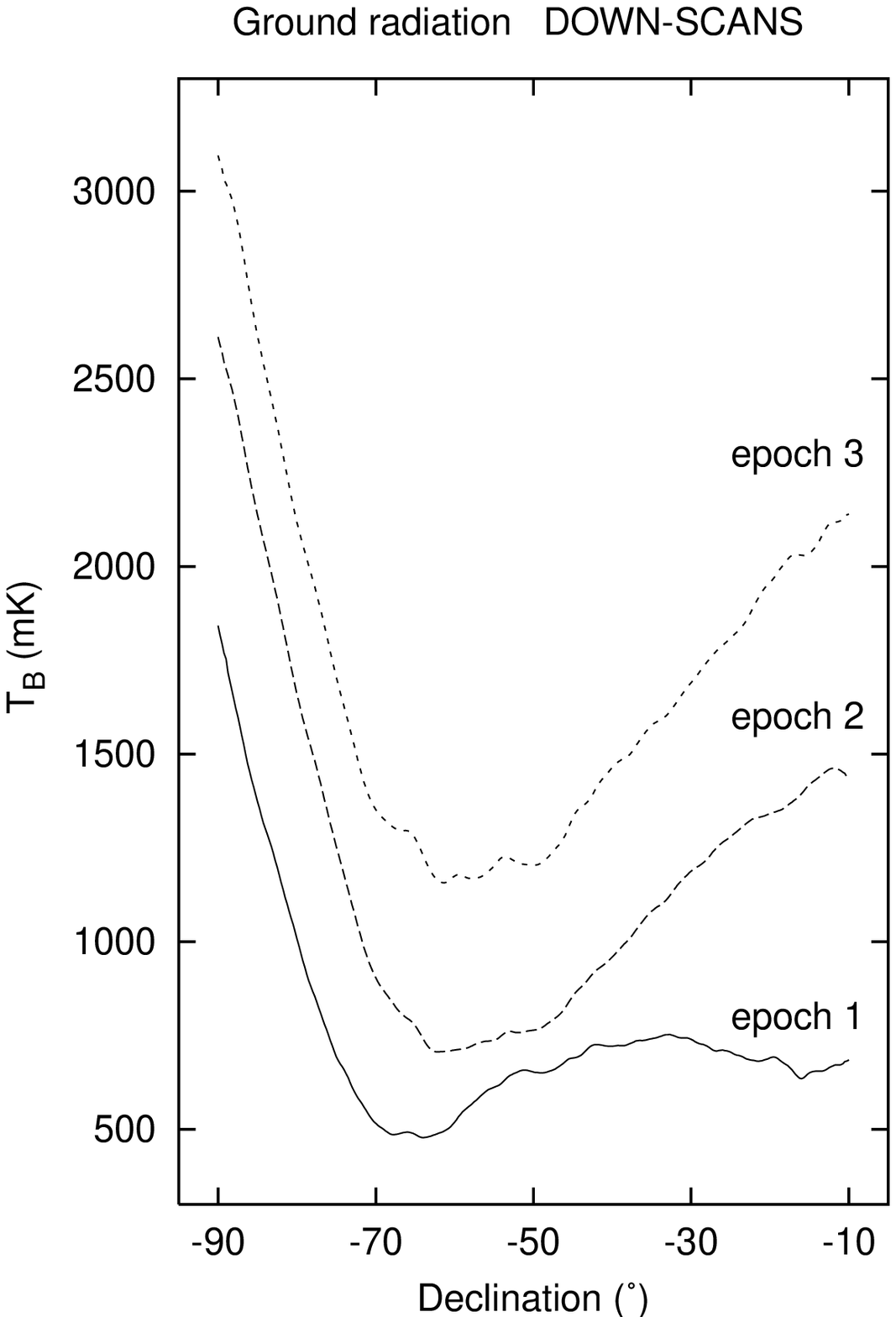}
}
\caption{Adopted ground radiation profiles for different epochs, which
were subtracted from the individual scans. The adopted zerolevels are
arbitrary. }
\label{fig_three}
\end{figure}

The tabular scan data were sorted in ascending right ascension,
calibrated, plotted and inspected and edited for distortions or
interference. By scanning in declination along the local meridian
the sidelobe structure and the ground radiation will be similar for all
scans, and therefore the contribution to the antenna temperature is
equal for all points at a given declination. Numerous scans observed in
regions of the sky where the emission is nearly constant ($\rm 22^h \leq
R.A. \leq 3^h$) were selected for the two different observing sequences
and a lower envelope fitted to these data to determine the
characteristic of the ground radiation. In the course of determining
the ground radiation curve for the observation period 1993/1994 it
became clear that this sequence had to be divided into two intervals.
Figure~\ref{fig_three} shows the ground radiation profiles for UP and
DOWN scans at the three different observation periods.

We have no quantitative explanation as to why the ground radiation
profiles are different for the three epochs nor why these curves show
a different behaviour for UP- and DOWN-scans in period 2\&3.
Small systematic changes of the system temperature depending on
elevation seem to be the most likely reason. These profiles were
subtracted from each observed scan before applying the baseline
correction procedure. In order to establish a consistent zero level for
the observations, the temperature scale and zero level were found by
comparison with the absolutely calibrated Stockert 1420-MHz survey in
the overlapping area between $-19\degr$ and $-10\degr$ in declination.
Each scan was corrected for the skew (`nodding') angle and precessed to
equatorial coordinates epoch 1950.0. In the common declination range
the mean temperature was calculated and compared with the Stockert
survey. In this way, we fixed the upper end of a scan. For the lower
end, we averaged the data below declinations of $-80\degr$ excluding
compact sources and set it to a constant temperature. With this data
set first raw maps were computed. The remaining ``scanning effects''
due to weather conditions and receiver instabilities were removed using
the so-called method of unsharp masking (Sofue \& Reich\
\cite{sofue+reich79}). We used an elliptical Gaussian beam of $1\fdg5
\times 0\fdg6$ as a smoothing function. First- and then second-order
polynomial fits to the difference temperatures from the smoothed data
were utilised to minimize the scanning effects. For the maps observed
during the second and third period we used the corrected  map from the
first observing period as a reference.

% Fig. 3

We ended up with four maps of the southern sky: two UP-scan and two
DOWN-scan. These maps were separately transformed to Epoch 1950. We
next computed the difference between the UP-scan map and the
DOWN-scan map of each epoch and that between the maps observed in the
same scanning direction but of the two different epochs in order to
check the pointing accuracy, the temperature scale and whether the
subtracted ground radiation profile was correct. We realized that the
bandstop filter used during the second coverage to exclude the
contribution from the local \ion{H}{i} emission was ineffective in
eliminating the \ion{H}{i} emission from the Small Magellanic Cloud
(SMC) and the Large Magellanic Cloud (LMC). Both areas were excluded
from the second coverage, i.e. the area covering $\rm 4^h 40^m \le R.A.
\le 6^h$ and $-75\degr \le \delta \le -60\degr$ and $\rm 0^h 20^m \le
R.A. \le 1^h 20^m$ and $-75\degr \le \delta \le -70\degr$. The
sensitivity was not reduced because the survey is oversampled in this
declination range.

The final map was obtained by adding the four maps with the
so-called PLAIT-algorithm (Emerson \& Gr\"ave\ \cite{emerson+graeve88}),
which in addition destripes the maps. The first coverage was given
a double weight.

The full-beam scaling was adopted from the Stockert survey (S/T$_{\rm
B}$ =~11.25~Jy/K), since no large-size antenna pattern could be
observed for the IAR 30-m telescope. While the obtained r.m.s.-noise
of the present data agrees with that of the Stockert northern sky
survey, we note that the southern sky maps show significantly less
scanning effects due to the more advanced software and processing
power available today.

%Fig. 4
\begin{figure*}[htbp]
\includegraphics[width=14truecm]{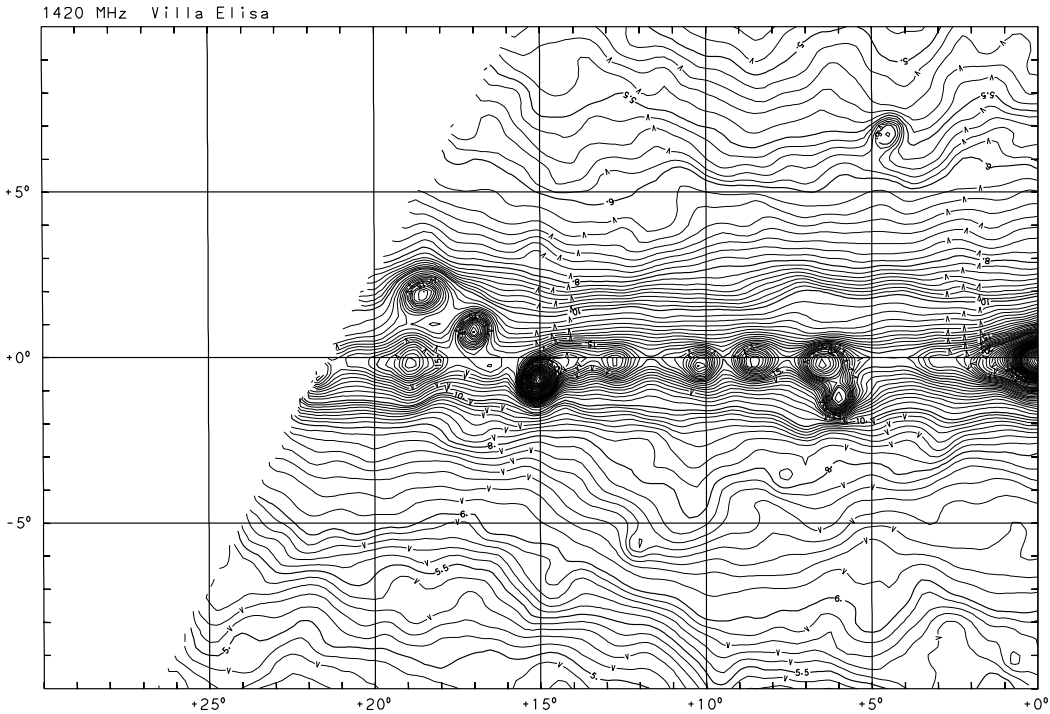}
\caption{Area towards the Galactic Centre as observed with the Villa
Elisa 30-m telescope. The contours are labelled in K $\rm T_{B}$ (full
beam). Contour steps are 100~mK up to 6~K, 250~mK up to 10~K, 500~mK up
to 20~K, 1~K up to 50~K and then 2.5~K.}
\label{fig_four}
\end{figure*}
%
%Fig. 5
\begin{figure*}[htbp]
\includegraphics[width=14truecm]{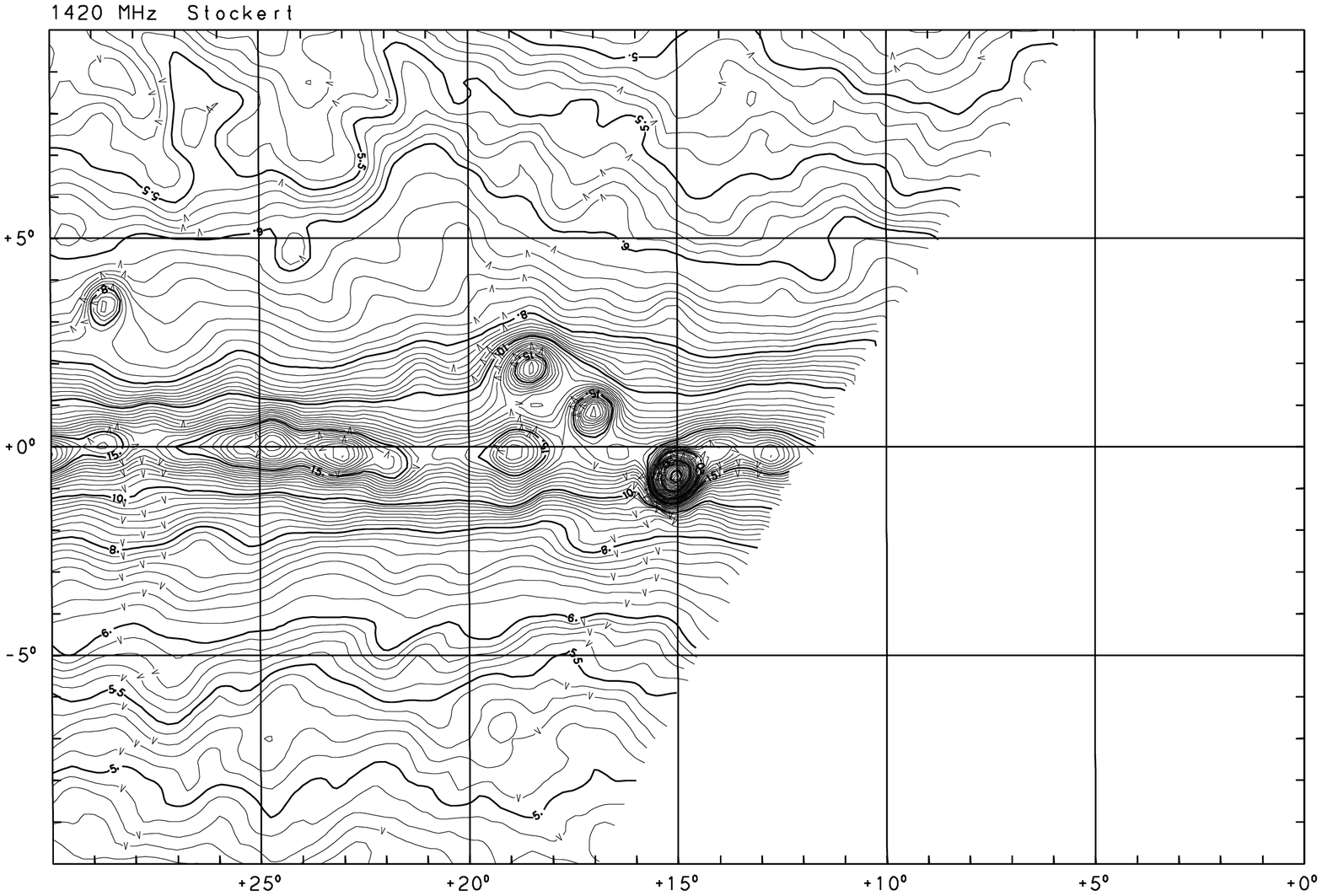}
\caption{Area towards the Galactic Centre as observed with the
Stockert 25-m telescope. The contour steps are the same as in
Fig.~\ref{fig_four}.}
\label{fig_five}
\end{figure*}

\section{Absolute calibration}

The Villa Elisa southern sky survey is tied to the Stockert northern
sky survey in the region of overlap, but any residual large-scale
temperature gradient towards the South Celestial Pole requires
the determination of absolute temperatures. Absolute temperatures have
been measured by Bensadoun et al. (\cite{bensadoun+93}), who have
performed sky-horn measurements at various declinations of the northern
but also of the southern sky by using the same equipment to determine
the cosmic microwave background temperature at 1.47~GHz. The resolution
of these data is $30\degr \times 27\degr$, to which we have convolved
both the data of the northern and the southern sky. For the northern
sky survey, which was absolutely calibrated by a comparison with data of
Howell \& Shakeshaft (\cite{howell+shake66}) and Pelyushenko \&
Stankevich (\cite{pely+stank69}) at a declination of $30\degr$, we
found a difference to the data of Bensadoun et al. (\cite{bensadoun+93})
of +0.53~K. The derived cosmic microwave background temperature by
Bensadoun et al. (\cite{bensadoun+93}) at 1.47~GHz of $2.26 \pm 0.19$~K
differs by 0.44~K to that assumed by Reich \& Reich
(\cite{reich+reich88a}) of 2.7~K. However, there is an additional
measurement of the cosmic microwave background of the northern sky at
1.4~GHz by Staggs et al. (\cite{staggs+96}). Their result is
$2.65^{+0.33}_{-0.30}$~K, rather close to the adopted temperature of
2.7~K for the northern sky. When subtracting a 2.7~K cosmic microwave
background contribution from the 1420~MHz data of the northern sky a
reliable spectral index map between 408~MHz and 1420~MHz (Reich \&
Reich\ \cite{reich+reich88a}) was obtained. Any residual temperature
offset at 1420~MHz must be small, e.g. not exceeding about 0.1~K,
otherwise a spectral change as a function of intensity results.

We take an offset of +0.49~K (e.g. the mean of both the differences
mentioned above) as the best compromise for a correction of the
Bensadoun et al. data and added this value to the measurements at
declination  $-75\degr$ and the equatorial south pole. We end up with
the following temperatures, averaged for the specified right ascensions,
of 3.30~K ($\rm 4^h 6^m - 6^h 0^m$) and 3.28~K ($\rm 3^h 42^m - 4^h
42^m$) for $\delta=-75\degr$ and 3.58~K at the south pole ($\rm 7^h
6^m - 7^h 42^m$). The temperatures from the Villa~Elisa 1420~MHz survey
are 3.59~K, 3.58~K and 3.59~K, respectively. Clearly, the temperatures
at the south pole agree quite well. However, the temperatures at
$-75\degr$ declination are about 0.3~K higher than the sky-horn data
and are nearly the same as measured at the south pole. We have at
present no explanation for this discrepancy compared to the sky-horn
data, although the rather steep ground radiation profiles
(Fig.~\ref{fig_six}) in this declination range might influence the
zero-level accuracy locally. There is, however, no indication of a
significant temperature gradient towards the South Celestial Pole
either from the 45~MHz survey of the southern sky (Alvarez et al.\
\cite{alvarez+97}) or the 408~MHz all-sky survey (Haslam et al.\
\cite{haslam+82}), which casts some doubt on the accuracy of the
sky-horn data.

\section{Example maps}

We show two example maps of the Villa Elisa southern sky survey in
comparison to the northern sky  Stockert data in their regions of
overlap. These maps clearly illustrate the comparable quality of the
data from both surveys as well as their common zero level and
temperature scale.

\begin{figure}[htb]
\includegraphics[width=\hsize]{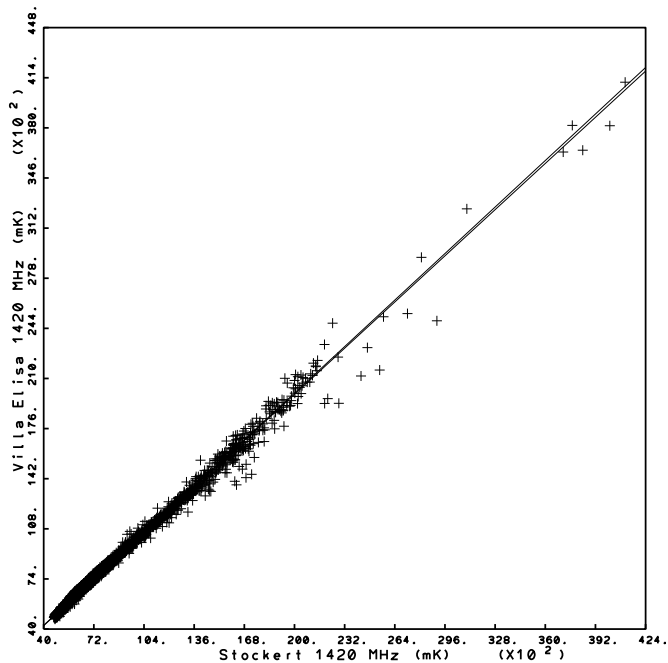}
\caption{T--T plot of the data shown in Figs.~\ref{fig_four} and
\ref{fig_five}. The slope of the linear fits is 0.979 or 1.014 for
fitting Villa Elisa data versus Stockert data and vice versa.}
\label{fig_six}
\end{figure}

Figures~\ref{fig_four} and \ref{fig_five} display a region of $30\degr
\times 20\degr$ along the Galactic plane north of the Galactic centre
area. The corresponding T--T plot (temperature versus temperature plot)
is shown in Fig.~\ref{fig_six}. In this region, rather high
temperatures were observed and these data are therefore well suited to
confirm the temperature scale of the two surveys. As seen from
Fig.~\ref{fig_six} both scales agree within about 2\%.

%Fig. 6

\begin{figure*}
\resizebox{14cm}{!}{\includegraphics{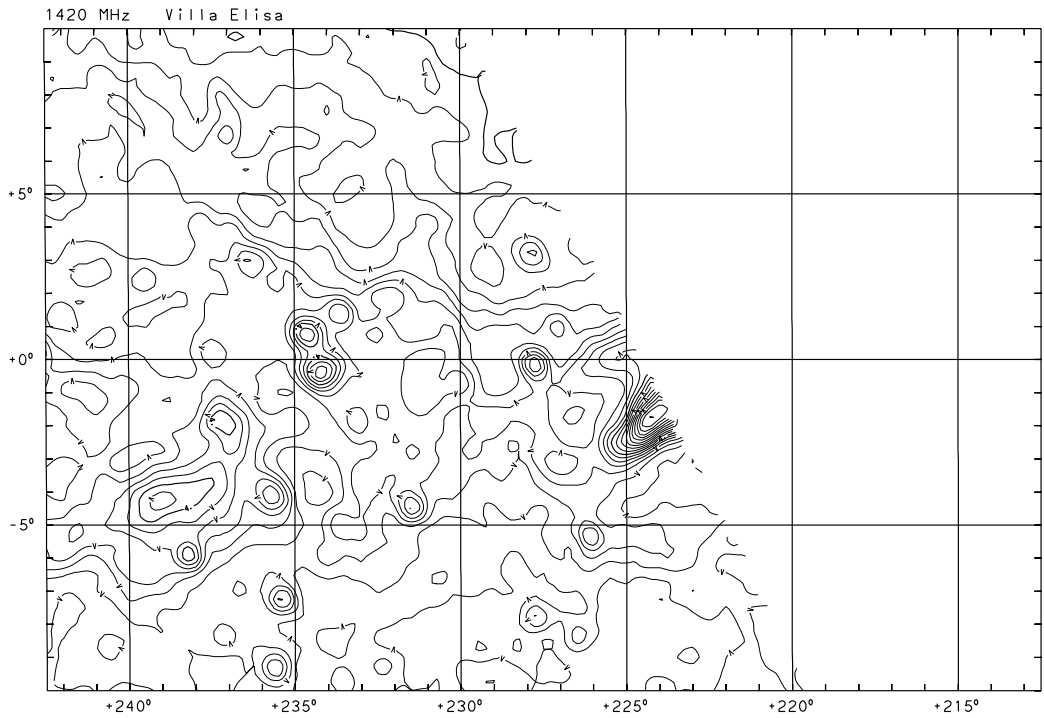}}
\caption{Area towards the Galactic Anti-centre as observed with the
Villa Elisa 30-m telescope. The contours are labelled in  K $\rm T_{B}$
(full beam) and the contour steps are 50~mK apart.}
\label{fig_seven}
\end {figure*}

\begin{figure*}
\resizebox{14cm}{!}{\includegraphics{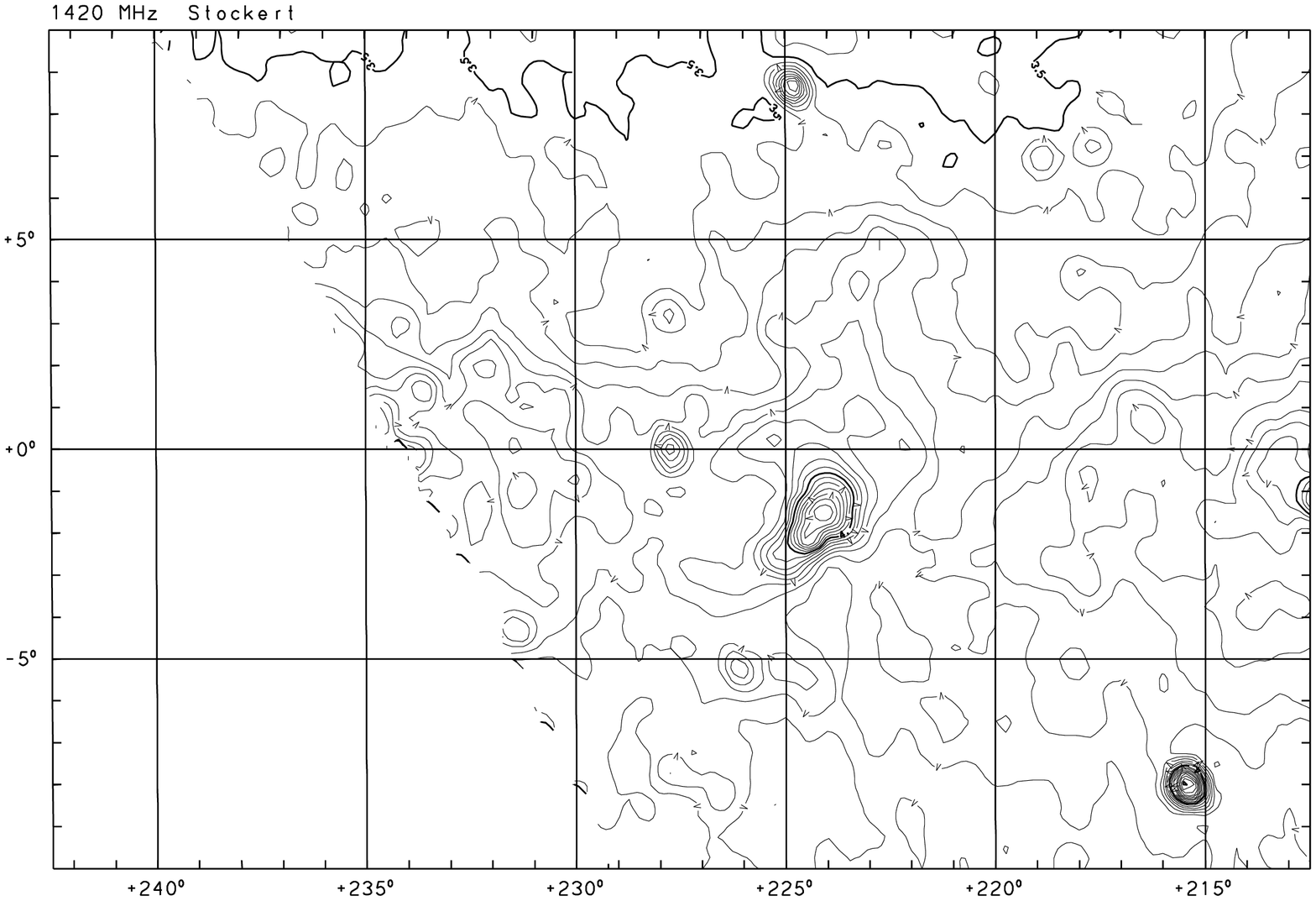}}
\caption{Area towards the Galactic Anti-centre as observed with the
Stockert 25-m telescope. The contour steps are the same as in
Fig.~\ref{fig_seven}.}
\label{fig_eight}
\end {figure*}

As a second example we show in Figs.~\ref{fig_seven} and
\ref{fig_eight} a section of the Galactic Anti-centre. The emission
level is rather low in this area, with just about 1~K above the cosmic
microwave background even in the Galactic plane. The two surveys
agree within one contour, which is 50~mK T$_{\rm B}$.

\begin{acknowledgements}
We are indebted to Prof. Richard Wiele\-binski for his support through
all stages of the survey project. We like to thank Dr. Glyn Haslam for
providing the on-line software for the receiver control, data
acquisition and formation of tabular scans.  We also acknowledge the
patience of Ursula Geisler for bookkeeping of the raw data of the
survey. J.C.T. thanks the Max-Planck-Gesellschaft for financial support
during his stay at the MPIfR. We would like to thank Dr.~Tom Wilson and
Dr.~Axel Jessner for comments on the manuscript.
\end{acknowledgements}

{}

% Fig. 7

\end{document}